\documentstyle[prl,aps,preprint]{revtex}

\begin{document}
\tighten

\title{PARTON-HADRON DUALITY:\\
RESONANCES AND HIGHER TWISTS \thanks
{This work is supported in part by funds provided by the U.S.
Department of Energy (D.O.E.) under cooperative agreement
\#DF-FC02-94ER40818.}}

\author{Xiangdong Ji and Peter Unrau}

\address{Center for Theoretical Physics \\
Laboratory for Nuclear Science \\
and Department of Physics \\
Massachusetts Institute of Technology \\
Cambridge, Massachusetts 02139 \\
{~}}

\date{MIT-CTP-2348  \hfill  HEP-PH/9408317  \hfill
Submitted to: {\it Phys.~Letts.~B}  \hfill  August 1994}

\maketitle

\begin{abstract}
We explore the physics of
the parton-hadron duality in the nucleon
structure functions appearing in lepton-nucleon scattering.
We stress that the duality allows one to extract
the higher-twist matrix elements from
data in the resonance region, and learn about the
properties of resonances if these matrix
elements are known. As an example, we
construct the moments of $F_2(x, Q^2)$
for the low and medium $Q^2$ region, and from which
we study the interplay between higher twists and the
resonance contributions.
\end{abstract}

\pacs{xxxxxx}

\narrowtext

In electron-nucleon scattering, one probes
the substructure of the nucleon with
virtual photons of mass $Q^2$ and energy $\nu$.
Before the advent of Quantum Chromodynamics (QCD),
Bloom and Gilman~\cite{BG} discovered
an interesting phenomenon about the
nucleon structure function $W_2(\nu, Q^2)$, measured at SLAC\null.
Simply speaking, when expressed in terms of the improved
scaling variable $\omega' = 1 + W^2/Q^2$,
where $W$ is the final-state hadron mass,
the scaling function $F_2(Q^2,\omega') = \nu W_2/m_N$
in the resonance region ($W<2$ GeV) roughly
averages to (or duals) that
in the deep-inelastic region ($W>2$ GeV).
Referring to a similar phenomenon observed
in hadron-hadron scattering, they called it
{\it parton-hadron duality}. Moreover,
the occurrence of the duality appears to be local,
in the sense that it exists for each interval of
$\omega'$ corresponding to the prominent
nucleon resonances. In fact, the assumption of
an exact local duality allows an approximate
extraction of
the nucleon's elastic form factor
from the deep-inelastic scaling function!

An explanation of the Bloom-Gilman duality in QCD
was offered by de Rujula, Georgi, and
Politzer in 1977~\cite{RGP}. Following the
operator product expansion, they studied the moments
of the scaling function in the Nachtmann scaling
variable $\xi = 2x/(1+\sqrt{1+4x^2m_N^2/Q^2})$,
where $x =Q^2/2m_N\nu$. They argued that
the $n$-th moment $M_n(Q^2)$ of $F_2$ has the following
twist expansion,
\begin{equation}
   M_n(Q^2) = \sum_{k=1}^\infty
\left({nM_0^2\over Q^2}\right)^{k-1}B_{n,k}(Q^2) \>,
\label{twist}
\end{equation}
where $M_0^2$ is a mass scale $\sim$ $(400\sim 500 {\rm MeV})^2$
and $B_{n, k}(Q^2)$ depends logarithmically on
$Q^2$ and is roughly on the order of $B_{n,0}$.
According to Eq.~(\ref{twist}), there exists
a region of $n$ and $Q^2$ ($n\le Q^2/M_0^2$),
where the higher twist contribution
is neither large nor negligible, and where
the dominant contribution to
the moments comes from the low-lying resonances.
The appearance of local duality reflects
the very existence of this region. A more recent
study on duality can be found in Ref.~\cite{CM}.

While these original studies of the parton-hadron duality
were largely qualitative,
enormous progress has been made in understanding QCD
in the past twenty years. The radiative corrections
have been evaluated to the next-to-leading order for
the twist-two part of the scaling function~\cite{BUR};
the structure of the higher twist expansion has been
clarified to the order of $1/Q^2$ and some at the order
of $1/Q^4$\cite{SV}. The physics of the parton-hadron duality
has been exploited ingeniously in the
vacuum correlation functions, from which
a powerful technique for calculating hadron properties
from QCD---the QCD sum rule method---has emerged~\cite{SVZ}.
Experimentally, a large body of lepton-nucleon
scattering data has been collected in the past 25 years~\cite{VIR}.
With the CEBAF facility becoming available
for making systematic, high precision measurements
in the resonance region, it
is timely to re-examine
duality in its original context,
and further explore the physics content of this
important concept.

In this Letter we seek to sharpen the explanation of
the duality offered by authors in Ref.~\cite{RGP},
with a few crucial differences. First, we choose to
work with the moments of Cornwell-Norton,
instead of those of Nachtmann, thereby avoiding the unphysical
region of $\xi>\xi(x=1)$. Second, we look for a
way to describe more clearly the contribution of the resonances
to the moments. Finally,
we emphasize a thorough exploitation of the
consequences of duality. We furnish our discussions with
the example of $F_2$, for which the abundant data
allow an accurate construction of its moments
in the low and medium $Q^2$ region. These moments
offer a unique opportunity for studying the
effects of higher twists and the resonance contributions.

The Cornwell-Norton moments of a scaling function $F(x, Q^2)$
are defined as,
\begin{equation}
       M_n(Q^2) = \int^1_0 dx x^{n-2} F(x, Q^2) \>,
\end{equation}
where the upper limit includes the elastic
contribution. According to the operator production expansion,
the moments can be expanded in powers of $1/Q^2$,
\begin{equation}
       M_n(Q^2) = \sum_{k=0}^{\infty}
           E_{nk}(Q^2/\mu^2)M_{nk}(\mu^2)
            \left({1\over Q^2}\right)^{k},
\label{expa}
\end{equation}
where $E_{nk}$ are the dimensionless coefficient functions
which can be calculated perturbatively as a power series
in the strong coupling constant $\alpha_s(Q^2)$,
\begin{equation}
         E_{nk}(Q^2/\mu^2) = \sum_{i=0}^\infty
        \alpha_s^i(Q^2) e^i_{nk} \>,
\label{pert}
\end{equation}
and $M_{nk}(\mu^2)$ are the nucleon matrix elements
of local operators composed of
quark and gluon fields. The renormalization scale
$(\mu^2)$ dependence cancels in the product of the two
quantities; however, when we talk about them
separately, $\mu^2$ is chosen to be the hadron mass scale.
The terms beyond the first in Eq.~(\ref{expa})
are called the higher-twist corrections, which include both
the target mass corrections and the true higher-twist
effects.

The double expansions in Eq.~(\ref{expa})
are asymptotic at best. Non-perturbative effects
can invalidate both expansions at higher orders, and
can mix the two, rendering the separation of
radiative and power corrections ambiguous~\cite{MUL}. In the
following discussion, however, we assume that in the
$Q^2$ region of our interest,
the size of the twist-four term ($1/Q^2$)
is significantly larger than the smallest term in the asymptotic
expansion for $E_{n0}$, beyond which
the evaluation of $E_{n0}$
cannot be improved by including higher-order terms, and
so the ambiguity in defining the higher-twist
corrections can be neglected~\cite{MUL}.
We shall henceforth focus only on the structure of the
twist expansion.

Following Ref.~\cite{RGP}, we
assume the ratio of the twist-four term to the
leading twist in each moment
is approximately $nM_0^2$,
where $M_0$ is a scale characterizing the matrix
elements of the twist-four operators.
We further assume that the twist expansion
is an asymptotic series in the
parameter $nM_0^2/Q^2$.
According to the above assumptions, we can
classify the higher-twist contributions to the moments.
Consider the $n-Q^2$ plane as shown in Fig.~\ref{fig1}(a),
which is separated into three regions by two solid lines.
Region A is defined by $n M^2_0 \ll Q^2$, where the
higher-twist effect are negligible. Region B
is where the higher-twist corrections become important
but stay perturbative, and thus
only the first few terms in the twist-expansion
are of practical importance. Region C
is where the higher-twist effects become
non-perturbative, and the power-expansion
loses meaning. It is in this
third region that the resonance physics
dominates the behavior of the moments
and the quantum coherence, inherent to
resonance production, defies a
description of the scattering in terms of
a finite number of quarks and gluons.
In the later part of the paper, we will
show that the first assumption is consistent
with the behavior of the lower moments for $F_2$.

Now we consider the resonance
contribution to the Cornwell-Norton moments by
examining the $x-Q^2$ plane shown in Fig.~\ref{fig1}(b),
in which the resonance region is approximately above the curve
$W=2$ GeV\null.
For a large, fixed $Q^2$ (say $15$ GeV$^2$),
the resonance contribution to the lowest few moments
is very small, and can be neglected.
When $n$ increases, the resonance contribution
is weighted more and becomes significant.
We can use a dashed line, as $Q^2$ varies,
in the $n-Q^2$ plane to indicate the separation of the
two cases. The
dashed line certainly cannot be in region A, because
the non-resonance experimental data have
already detected the higher-twist effects~\cite{VM}. If the dashed line
is in the region C, then the perturbative higher-twist
effects have nothing to do with resonance physics.
The most exciting possibility is when the dashed line
lies in the region B, and this is what happened in reality.

When the dashed line is located in
region B, then in the left portion of it,
the following statements are true:
1). the higher-twist corrections are
perturbative, so the moments are not too different from those
at larger $Q^2$, and 2). the resonance contribution to the
moments are significant. Thus in this region,
the resonances must organize themselves to
follow the deep-inelastic contribution apart
from a perturbative higher-twist correction,
or conversely, the structure of the higher-twist expansion
constrains the behavior of the resonance contribution.
The degree of duality is determined by the size of this
region: the larger the region, the more the moments
are constrained, and the more local the duality will~be.

Why should duality occur at all in QCD?
On one hand, the quark transverse momentum in
the nucleon, which governs the magnitude of
the higher-twists, is about 400 MeV\null.
This makes the higher-twist corrections
perturbative down to very small $Q^2$.
On the other hand, the resonance
contribution to the moments are already
significant at $Q^2 \sim 5 {\rm GeV}^2$ for
low $n$. Thus the occurrence of the duality seems
unavoidable, unless QCD had two widely
different~scales.

The consequences of duality, like duality itself,
are two-fold. If one knows data in the resonance region,
one can extract the matrix elements of the
higher-twist operators.
The extraction, of course, is limited
by our ability to calculate higher-order
radiative corrections, about which we have nothing to say.
On the other hand, if one knows the higher-twist matrix
elements from other sources, such as lattice QCD calculations,
one can utilize them to
extract the properties of the resonances.
This second use of duality
has been pursued extensively in the QCD
sum rule calculations, from which a large number of interesting results
has been obtained~\cite{SVZ}. In the present case, however,
the number of higher-twist matrix elements is
large, and they are difficult to estimate in general.
This severely limits our ability to check, for example,
the internal consistency of the duality predictions.

We make the above discussion more concrete
and quantitative by using the example of
the $F_2$ scaling function, for which rich
data exist in an extended kinematic region.
Most of the low $Q^2$ data were
taken in late 60's and early 70's at SLAC and DESY,
and they nearly cover the whole resonance region
at large $x$. The data were fitted by Brasse {\it et al}.~\cite{BRASS}
to a function with three parameters for each fixed $W$.
In Ref.~\cite{BODEK}, Bodek {\it et al}.\
have made a more extensive but different fit,
covering higher $Q^2$ resonance data.
The deep-inelastic data
were systematically taken by SLAC, BCDMS, EMC, and other
collaborations during the 70's and 80's, and they
have recently been shown to be consistent with
each other~\cite{VM}.
New measurements from NMC at CERN has extended
these data to lower $Q^2$ and $x$~\cite{NMC1}. In Fig.~\ref{fig2},
we have shown the $F_2$ data as a function of
Bjorken $x$ at $Q^2$ = 0.5, 1.0, 2.0, 4.0, 8.0
and 16.0 GeV$^2$ from the two fits \cite{BODEK,NMC1}
made in different kinematic regions.

The salient features of the data can be summarized as
follows. At high-$Q^2$, the data is almost entirely
deep-inelastic except for a small resonance contribution
at large $x$. The scaling function near
$x=0$ shows a rise due to perturbative QCD effects.
As $Q^2$ decreases, small bumps become visible and
slide toward low $x$.  These prominent excitations are
believed to be the $\Delta(1232)$,
$S_{11}(1535)$ or $D_{13}(1520)$, and
$F_{15}(1680)$ resonances. The resonance excitations
become very strong near $Q^2=2$ GeV$^2$
and clearly dominates $F_2$ below $Q^2=1$ GeV$^2$.
As $Q^2\rightarrow 0$ the data is compressed toward
$x=0$ due to simple kinematics. At $Q^2=0$,
the whole photo-production physics is shrunk to $x=0$. Of course,
one should not forget about the elastic contribution, which
contributes a delta-function at $x=1$.

To understand the role of the resonances
in the Cornwell-Norton moments, we plot
in Fig.~\ref{fig3} the ratio of the resonance part
to the total, where the resonance contribution is defined
by a cut on $W<$ 2 GeV\null. If one uses ten
percent as a measure of the importance of the
resonance contribution, then this threshold
is reached for the lowest moment($n=2$) at
$Q^2 \sim 4 {\rm GeV}^2$.  For higher moments,
the transition occurs approximately at $2n$ GeV$^2$.
This is quite surprising because the non-perturbative
physics becomes potentially important
at $Q^2=16$ GeV$^2$ for the 8th moment! At
$Q^2=8$ GeV$^2$, the same moment receives fifty
percent of the contribution from the resonance
region. The dashed line
in Fig.~\ref{fig1}(a) roughly corresponds to the ten percent line shown
in Fig.~\ref{fig3}.

The data on the $F_2$ moments can be
used to extract the matrix
elements of higher-twist operators. To effect this,
we first subtract the twist-two part of the contribution.
We use a parton distribution (CTEQ2,~\cite{CTEQ})
fitted to a large number of data on hard
processes, and calculate the moments for each quark
flavor and gluon distribution at some large $Q^2$
(=20 GeV$^2$ in our case).
Then we evolve these moments to lower $Q^2$
using the perturbative QCD formula accurate
to next-to-leading order. Theoretical errors
in evolution are mainly generated from uncertainty in
$\Lambda_{\rm QCD}$ and
unknown higher-order terms in the coefficient functions.
In our work, we take $\Lambda_{\rm QCD}^{(4)} = 260 \pm 50$
GeV~\cite{PD},
and the resulting error is added to the experimental error
which is taken to be 3\% uniformly, yielding the total
error on the residue. The target mass corrections are
further subtracted from the moments according to
the formula in Ref.~\cite{GP}.
In Fig.~\ref{fig4}(a), we show the moments as a
function of $Q^2$ and the twist-two part
plus the target mass corrections (solid lines).
The residual moments, which are entirely higher
twist effects, are shown in Fig.~\ref{fig4}(b)
as functions of $1/Q^2$.

We choose to fit the $Q^2$ evolution of the moments with a
pure twist-four contribution,
\begin{equation}
    \Delta M_n(Q^2) = a_n \left({\alpha(Q^2)\over
     \alpha(1)}\right)^{\gamma_n} {1\over Q^2}
\end{equation}
where we have included phenomenologically the
leading-log effects with an adjustable exponent.
The fitted $\gamma_n$ represents an average of the
anomalous dimensions of the spin-$n$, twist-four
operators, weighted by the size of individual
matrix element. The coefficient $a_n$ is a simple
sum of the twist-four matrix elements at the scale
$\mu^2=1$ GeV$^2$. Inclusion of a twist-six term
creates strong correlations among the parameters and
renders the fits indeterminate. Thus we have
neglected such a term by restricting the fit to
the region with $Q^2>n$, where the twist-six contribution
is presumably small.

The result of our fit is shown in Table~\ref{tab1}. The correction to
the $n=2$ moment (the famous momentum sum rule)
is best determined, yielding a
characteristic higher-twist scale of 500 MeV\null.
{}From this, we determine that the twist-four
contribution to the momentum sum rule at
$Q^2=2$ GeV$^2$ is 0.015, about ten percent of the total.
The exponent of the leading-log contribution
increases gradually with $n$, in accord with general expectations.
The near constancy of the twist-four contribution
is in sharp contrast with the fast decrease of the
leading-twist contribution with increasing n.
It confirms, though, the speculation that the higher-twist
contribution become more important for
higher moments, and is a precursor for
the onset of the resonance region. In QCD, this
can be explained by an increasing number of
twist-four operators compensated by a decrease in strength of
individual matrix elements.
The pattern of the moments indicates a twist-four distribution
negative at small $x$, positive at large $x$
and peaked near $x=1$, qualitatively consistent with
the fits in Ref.~\cite{VM}, where
the resonance data were entirely ignored.

Finally, we test the assumption about the higher-twist matrix
elements in Eq.~(\ref{twist}). We show in the fourth column of
Table~\ref{tab1} the ratio
of the higher-twist matrix elements
and the twist-two part. From this, we
extract an effective $M_0$ by dividing by $n$
and taking the square root. The result is shown in the
fifth column and is approximately $n$-independent,
although there is a slight
hint of $M_0$ getting larger for larger $n$. However, this
should not be taken too seriously because of the errors and limited
number of moments. If fifty percent of the higher-twist
contribution is taken as an indication that the twist-expansion is
getting non-perturbative, we find a $Q^2$ for each moment
where the transition takes place.
For $n=2$, this is about 0.3 GeV$^2$.
For higher moments, this happens at about $n-1$ GeV$^2$.
The line which separates regions B and C in Fig.~\ref{fig1}
roughly corresponds to
this. Thus the existence of the duality zone
is clearly established beyond any doubt.

To illustrate the other use of duality, one
could, for instance, use the higher-twist contribution
extracted from the pure deep-inelastic region (as done in~\cite{VM}),
or from some theoretical calculations, to determine
the nucleon's elastic form factor.
However, we feel that the higher-twist matrix elements
have not been determined in other methods
to a sufficient accuracy to allow a quantitative
extraction of the resonance properties.
Qualitatively, however, knowing the higher-twist contribution
will surely improve the nucleon form factor
extracted in Ref.~\cite{RGP}, which shows a systematic
deviation from the directly measured $G_M$,
a clear indication of higher-twist effects.

To sum up, we explored in this work the physics
of the parton-hadron duality. We emphasized that
the existence of duality allows one to determine
the higher twist matrix elements from data in the
resonance region, or alternatively, knowing the matrix
elements enables one to determine the
properties of the resonances.
We studied the duality picture offered by the
$F_2$ scaling function, and extracted the
matrix elements of the lowest few spin, twist-four operators.
Clearly, this study can be applied straightforwardly
to the spin-dependent
structure function $G_1$ once more data becomes
available.

\acknowledgments
We thank A. Bodek, J. Morfin, and P. Stoler for
information on the experimental data and the CETQ2
parton distributions. PU acknowledges the support
of a NSERC scholarship from the
Canadian government.

\begin{figure}
\bigskip
\caption{a).~~Three regions of differing importance to higher twists:
Region A, negligible higher twists; Region B, perturbative higher
twists; and region C, the twist-expansion breaks down.
{}~b).~~Kinematic
regions corresponding to the resonance and deep-inelastic scattering.}
\label{fig1}

\bigskip
\caption{Scaling function obtained from the fits to experimental data
in Refs.~\protect\cite{BODEK,NMC1}.}
\label{fig2}

\bigskip
\caption{Ratio of the moments from the resonance region, including
the elastic contribution, to that of the total.}
\label{fig3}

\bigskip
\caption{a).~~Moments as functions of $Q^2$,
extracted from the scaling function in Fig.~\protect\ref{fig2}.
The solid lines refer to the contribution
from the leading twist and target-mass corrections.
{}~b).~~Residue moments from the higher-twist contribution.  The solid lines
are the fits
described in the text.}
\label{fig4}
\end{figure}

\begin{table}
\bigskip
\centering
\caption{Extracted twist-four matrix elements $a_n$,
effective anomalous dimension $\gamma_n$, ratio to the leading
twist contribution, and the effective mass scale $M_0$.}
\medskip
\label{tab1}
\mediumtext
\begin{tabular}{|@{\hspace{2em}}c@{\hspace{2em}}|c@{\hspace{2em}}|
           c@{\hspace{2em}}|d@{\hspace{3em}}|d@{\hspace{3em}}|}
$n$ & $a_n({\rm GeV}^2)$ & $\gamma_n$ & $a_n/(E_{n0}M_{n0})$ & $M_0$ \\ \hline
2 & $0.030\pm0.003$ & $1.0\pm0.5$ & 0.14 & 0.26 \\
4 & $0.042\pm0.013$ & $1.5\pm0.5$ & 1.00 & 0.50 \\
6 & $0.047\pm0.021$ & $2.5\pm0.5$ & 2.47 & 0.64 \\
8 & $0.038\pm0.018$ & $2.5\pm0.5$ & 3.45 & 0.66 \\
10 & $0.052\pm0.025$ &$ 3.5\pm0.5$ & 4.73 & 0.69 \\
\end{tabular}
\end{table}

\end{document}